\documentclass[aps,prd,twocolumn,showpacs,eqsecnum,amsmath,nofootinbib]{revtex4}
\usepackage[T1]{fontenc}
\usepackage[utf8]{inputenc}
\usepackage{lmodern}
\usepackage[english]{babel}

\usepackage{graphicx,amssymb,amsthm,amsmath}
\usepackage{hyperref}

\def \d {{\rm d}}
\def \e {e}

\def \boldl {\mbox{\boldmath$l$}}

\begin{document}

\title{Properties of Robinson--Trautman solution with scalar 
hair}

\author{T. Tahamtan}
\email{tahamtan@utf.mff.cuni.cz}

\author{O. Sv\'{\i}tek}
\email{ota@matfyz.cz}
\affiliation{Institute of Theoretical Physics, Faculty of Mathematics and 
Physics, Charles University in Prague, V~Hole\v{s}ovi\v{c}k\'ach 2, 180~00 
Prague 8, Czech Republic}

\begin{abstract}
An explicit Robinson--Trautman solution with minimally coupled free scalar field 
was derived and analyzed recently. It was shown that this solution possesses 
a curvature singularity which is initially naked but later enveloped by a horizon. However, this study concentrated on the general branch of the solution where 
all free constants are nonzero. Interesting special cases arise when some of 
the parameters are set to zero. In most of these cases the scalar field is still 
present. One of the cases is a static solution which represents a parametric 
limit of the Janis--Newman--Winicour scalar field spacetime. Additionally, we provide a calculation of the Bondi mass which clarifies the interpretation of the general solution. Finally, by a complex rotation of a parameter describing the strength of the scalar field we obtain a dynamical wormhole solution.
\end{abstract}

\date{\today}

\maketitle

\section{Introduction}
The Robinson--Trautman family of spacetimes contains nonsymmetric dynamical generalizations of several important solutions to the Einstein field equations --- e.g., Schwarzschild and Vaydia solutions or the C-metric. Notably, these solutions generally contain gravitational waves and offer the possibility to study the evolution of initial generic data towards a final stationary situation. This family is defined by the presence of a nontwisting, nonshearing, and expanding null geodesic congruence.

Recently, we presented a Robinson--Trautman solution minimally coupled to a free massless scalar field \cite{Tahamtan-PRD-2015} (a broader overview of the standard Robinson-Trautman solutions with many references can be found there). In this case, it was not 
possible to use the original form of the Robinson-Trautman metric which admits only pure radiation and the Maxwell field stress 
energy tensor aligned with the principal null direction or a cosmological constant. The reason was that the 
scalar field wave equation can not be satisfied for a scalar field whose gradient 
is aligned. The scalar field had to become non-aligned and the 
Robinson--Trautman metric had to be generalized to accommodate a broader class 
of energy-momentum tensors. Complete classification of such geometries including 
general forms of the curvature tensors is shown in \cite{Podolsky-Svarc}.

To help us with the interpretation of our solution we compute the Bondi mass \cite{Bondi} which is the most suitable description of the energy content of the spacetime for the Robinson-Trautman class and was used in this context previously (see, e.g., \cite{Tafel} for the vacuum case computation by means of a conformal factor). We use the definition based on an asymptotic twistor equation adapted to a massless scalar field in \cite{Scholtz}. The computation confirms the expectation based on an asymptotic form of the solution. Namely, the energy content of the spacetime is completely determined by a scalar field and there is no contribution of a Schwarzschild-type ``mass''. 

Next, we show the specific subcases which are all spherically symmetric and we connect them with previously analyzed solutions. This brings a certain degree of justification for considering the whole class as physically relevant.

Finally, inspired by a relation between a static subcase of our solution and a simple wormhole spacetime we analyze a wormhole version of the general Robinson--Trautman solution with a scalar field. Unlike the original solution, the wormhole version naturally does not satisfy any energy conditions and thus is not very physical. On the other hand, it provides a dynamical wormhole which gets created and then disappears while having surprising asymptotic behavior which is connected to the Kundt class. This family of solutions to the Einstein equations is closely related to the Robinson--Trautman geometry and is defined by the presence of a nontwisting, nonshearing, but (unlike Robinson--Trautman) nonexpanding null geodesic congruence. Most important members of this family are exact radiative spacetimes which generalize simple planar gravitational waves (e.g. pp-waves).

\section{Vacuum Robinson--Trautman metric and field equation}
\label{RTmetricsec}
First, let us review the standard Robinson--Trautman solution for comparison and reference.

The vacuum Robinson--Trautman spacetime (possibly with a cosmological constant $\Lambda$) can be described by the line element \cite{RobinsonTrautman:1960, RobinsonTrautman:1962, Stephanietal:book, GriffithsPodolsky:book}
\begin{equation}\label{RTmetric}
\d s^2 = -2H\,\d u^2-\,2\,\d u\,\d r + \frac{r^2}{\mathcal{P}^2}\,(\d x^{2} + \d y^{2}),
\end{equation}
where ${2H = \Delta(\,\ln \mathcal{P}) -2r(\,\ln \mathcal{P})_{,u} -{2m/r} -(\Lambda/3) r^2}$,
\begin{equation}\label{Laplace}
\Delta\equiv \mathcal{P}^2(\partial_{xx}+\partial_{yy}).
\end{equation}
The metric generally contains two functions, ${\,\mathcal{P}(u,x,y)\,}$ and ${\,m(u)\,}$. The function $m(u)$ might be set to a constant by suitable coordinate transformation \cite{Stephanietal:book, GriffithsPodolsky:book} and we consider this to be fulfilled for the coordinates of (\ref{RTmetric}). The Einstein equations then reduce to a single nonlinear PDE --- the Robinson--Trautman equation
\begin{equation}
\Delta\Delta(\,\ln \mathcal{P})+12\,m(\,\ln \mathcal{P})_{,u}=0\,.
\label{RTeq}
\end{equation}
These spacetimes are then generally of algebraic type II.

As required by the definition of the Robinson--Trautman family the spacetime admits a geodesic, shearfree, twistfree and expanding null congruence generated by ${\boldl=\partial_r}$ with $r$ being an affine parameter along this congruence, $u$~is a retarded time and $u=const$ hypersurfaces are null. Spatial coordinates $x,y$ span a transversal 2-space which has the Gaussian curvature (for ${r=1}$) 
\begin{equation}
\mathcal{K}(x,y,u)\equiv\Delta(\,\ln \mathcal{P})\,.
\label{RTGausscurvature}
\end{equation}
For general $r=const$ and $u=const$, the Gaussian curvature is ${\mathcal{K}/r^2}$ so that, as ${r\to\infty}$, these 2-spaces become locally flat. As usual we will assume that the transversal 2-spaces are compact and connected which leads to a subclass that contains the Schwarzschild solution (considering vanishing cosmological constant for simplicity) corresponding to $\mathcal{K}=1$ (consistent with a spherical symmetry). This subclass thus represents its generalization to a nonsymmetric dynamical situation.

For analysis of the Robinson--Trautman equation \eqref{RTeq} it is useful to introduce the following parametrization
\begin{equation}\label{PfP0}
\mathcal{P}=f(x,y,u)\, \mathcal{P}_{0}\,,
\end{equation}
where $f$ is a function on a 2-sphere $S^{2}$, corresponding to $\mathcal{P}_0=1+\frac{1}{4}(x^2+y^2)$ (such choice gives $\mathcal{K}=1$). By a rigorous analysis of the equation (\ref{RTeq}) together with the decomposition (\ref{PfP0}) Chru\'sciel \cite{Chru1,Chru2} proved that, for an arbitrary, sufficiently smooth initial data ${f(x,y,u_\textrm{i})}$ on an initial hypersurface ${u=u_\textrm{i}}$, the Robinson--Trautman type~II vacuum spacetimes \eqref{RTmetric} exist globally for all ${u\ge u_\textrm{i}}$. Moreover, they asymptotically converge to the Schwarzschild--(anti-)de~Sitter metric with the corresponding mass $m$ and the cosmological constant $\Lambda$ as ${u\to+\infty}$. This convergence is exponentially fast since $f$ behaves asymptotically as
\begin{equation}
f= \sum_{i,j\ge0} f_{i,j} u^j \e^{-2iu/m}
\end{equation}
where $f_{i,j}$ are smooth functions of the spatial coordinates $x,y$. For large retarded times~$u$, the function $\mathcal{P}$ given by \eqref{PfP0} exponentially approaches ${\mathcal{P}_0}$ which describes the corresponding spherically symmetric solution.

There is a closely related family of exact solutions possessing a null congruence which is nonexpanding (unlike in the case of Robinson--Trautman class), but still nontwisting and nonshearing --- namely the Kundt family \cite{Kundt1,Kundt2}. The general line element can be given in the form \cite{Stephanietal:book, GriffithsPodolsky:book} (we use the same coordinate labels as in the Robinson--Trautman case to stress the similarities and differences)
\begin{eqnarray}\label{Kundt-metric}
 \d s^2&=&-H\,\d u^2-2\,\d u\d r -2\,W_{1}\,\d u\d x\\
 && - 2\,W_{2}\,\d u\d y+\frac{\d x^2+\d y^2}{P(u,x,y)^2}\nonumber
\end{eqnarray}
with $H, W_{1}, W_{2}$ being functions of all the coordinates. Note that the coordinate $r$ is absent from the transversal part of this metric which is given by the last two terms of (\ref{Kundt-metric}) (compare with the Robinson--Trautman metric (\ref{RTmetric})). This directly leads to the privileged null congruence $\partial_{r}$ being nonexpanding. As we will see in the section \ref{wormhole} this solution is closely related to an asymptotic state of the Robinson--Trautman geometry with an imaginary scalar field. 

\section{Robinson--Trautman solution coupled to a scalar field}
The Robinson--Trautman solution generalized to accommodate a scalar field is given in the following form \cite{Tahamtan-PRD-2015}
\begin{eqnarray}\label{ourmetric}
 \d s^2&=&-2(h(u,r)+K(u,x,y))\,\d u^2-2\,\d u\d r \nonumber\\
&& +\frac{R(u,r)^2}{P(x,y)^2}(\d x^2+\d y^2)\ .
 \end{eqnarray}
The scalar field is assumed to be a function of $u$ and $r$ only ($\varphi(u,r)$). 
The dependence on $r$ means that the scalar field is not aligned (with respect to the direction given by its gradient). From the Einstein 
equations and the field equation for the scalar field
\begin{equation}\label{box}
\Box \varphi(u,r)=0
\end{equation}
(where $\Box$ is a standard d'Alembert operator for our metric (\ref{ourmetric}))
we obtained \cite{Tahamtan-PRD-2015} the following expressions for unknown 
metric functions and the scalar field 
\begin{eqnarray}\label{solution}
h(u,r)&=&\frac{r}{2U(u)}\frac{\partial U(u)}{\partial u} \nonumber\\
R(u,r)&=&\sqrt{\frac{U(u)^{2}r^2-C_{0}^2}{U(u)}}  \nonumber\\
K(u,x,y)&=&\frac{k(x,y)}{2U(u)} \\
k(x,y)&=&\Delta(\,\ln P(x,y))\nonumber\\
\Delta k(x,y)&=&{\alpha^2} \nonumber\\
U(u)&=&\gamma e^{\omega^2 u^2+\eta u}\nonumber\\
\varphi(u,r)&=&\frac{1}{\sqrt{2}}\ln{\left\{\frac{U(u)r-C_{0}}{U(u)r+C_{0}}\right \}}\nonumber
\end{eqnarray}
with this constraint between the constants appearing above
\begin{equation}\label{constraint}
 \omega=\frac{\alpha}{2C_{0}}\ .
\end{equation}

Unlike the case of the vacuum Robinson--Trautman spacetime (\ref{RTmetric}) one needs to solve three Einstein equations with nontrivial right-hand side given by the stress energy tensor of the scalar field. The relation corresponding to the original Robinson--Trautman equation (\ref{RTeq}) is now transformed into
\begin{eqnarray}\label{Euu}
	2\left(2\frac{R_{,r}}{R}H_{,r}+H_{,rr}\right)(H+K)+2\frac{R_{,r}}{R}(H+K)_{,u}&&\nonumber\\
	-\frac{2}{R}(R_{,u}H_{,r}+R_{,uu})+\frac{P^2}{R^2}(K_{,xx}+K_{,yy})=\varphi_{,u}^2\ .&&
\end{eqnarray}
One can recognize double Laplacian in the last term on the left-hand side (note the expression for $K$ in (\ref{solution})) and the second term of (\ref{RTeq}) has its analog at the beginning of the second line of (\ref{Euu}). The main reason for the difference between the vacuum (\ref{RTmetric}) and the scalar field (\ref{ourmetric}) case is an incompatible separation of variables for the metric function standing in front of the spatial part $\d x^2+\d y^2$. In the vacuum case the dependence on coordinates is split into $\{r\}$ and $\{u,x,y\}$ while in the scalar field case it is $\{u,r\}$ and $\{x,y\}$. Additionally, the scalar field case has more nontrivial Einstein equations and also the scalar field equation to satisfy. This led to a solution which is more explicit rather than being left unintegrated as usual for the vacuum Robinson--Trautman metric where it is possible to prove the existence of a solution for the single Robinson--Trautman PDE (\ref{RTeq}). Although both the vacuum (even with pure radiation) and the scalar field Robinson--Trautman solutions are of algebraic type II (see Appendix) the Ricci (or Segre) type is different and most importantly the scalar field is not aligned with the principle direction $\partial_{{r}}$ of the Weyl tensor.

In the previous study \cite{Tahamtan-PRD-2015} only a situation in which 
the constants $ C_{0}, \alpha, \eta, \gamma, \omega $  satisfy $ C_{0} > 0, \alpha > 0, \eta > 0, \gamma > 0$ was studied and the position of a curvature singularity and the existence of horizons was analyzed. Namely, it was shown that the singularity seems initially naked and only later it gets covered by the horizon. The asymptotic behavior of the metric was only hinted at in the original paper and since it shares its form with the special cases investigated later we will first derive the metric form when $u \to \infty$.

\section{Asymptotic behavior}\label{section-asymptotic}
To arrive at the asymptotic form of the general metric \ref{ourmetric} when $u$ goes to infinity one first notes that 
\begin{equation}\label{}
R(u,r) \simeq r\sqrt{U(u)}
\end{equation}
for large values of $u$ when parameter $\omega$ is nonzero in the solution (\ref{solution}). Combining functions $P$ and $U$ together
\begin{equation}\label{P-separated}
\mathcal{P}(u,x,y)=\frac{P(x,y)}{\sqrt{U(u)}}
\end{equation}
it is easy to see that 
\[K(u,x,y)=\frac{k(x,y)}{2U(u)}=\frac{\Delta(\,\ln \mathcal{P})}{2}\]
where  $k(x,y)=\Delta(\,\ln{P})={P}^2(\partial_{xx}+\partial_{yy})\, \ln{P}$.
Moreover, noting that
\[ \mathcal{P}_{,u}=-\frac{P}{2\,\sqrt{U}}(\,\ln U)_{,u}\]
one can express another metric function in a familiar form
\[h(u,r)=\frac{r}{2}(\,\ln U)_{,u}=-r(\,\ln \mathcal{P})_{,u} \]
Collecting these pieces together we arrive at the metric 
\begin{eqnarray}
\d s^2 &=& -[\Delta(\,\ln \mathcal{P})-2\,r(\,\ln \mathcal{P})_{,u}]\,\d 
u^2-\,2\,\d u\,\d r\nonumber\\
&&+ \frac{r^2}{\mathcal{P}^2}\,(\d y^{2} + \d x^{2}),
\end{eqnarray}
which is exactly of the original Robinson--Trautman form (\ref{RTmetric}) when $m=0$ and $\Lambda=0$. So in the asymptotic region the generalized form of the Robinson--Trautman metric evolves into a standard one with vanishing mass parameter and without cosmological constant. Such solutions belong to algebraic type N subclass of the original Robinson--Trautman spacetimes \cite{Stephanietal:book, GriffithsPodolsky:book} and contain (apart from a trivial flat solution) radiative solutions possessing singularity for certain values of $x,y$ on each wave-surface. These combine into singular lines in the spacetime. As noted in \cite{Tahamtan-PRD-2015} and repeated in appendix, the scalar field family of the Robinson--Trautman type does not contain these undesirable nontrivial type N solutions and therefore the final asymptotic state is just a flat spacetime (consistent with the vanishing scalar field in the asymptotic region). The specific reason for the absence of type N solutions with singular lines is the separated form of $\mathcal{P}$ (\ref{P-separated}) which is the result of the selected form of the metric (\ref{ourmetric}). 

Note, that the approach to the final asymptotic state differs from the vacuum case. In the vacuum case the behavior is described by a simple exponential (\ref{RTeq}) while for the scalar field case the exponential depends quadratically on the retarded time (see (\ref{P-separated}) and (\ref{solution})) provided $\omega\neq 0$. In this case the solution with a scalar field can additionally be extended to a negative infinite retarded time unlike in the vacuum case where the evolution generally blows up due to the parabolic nature of (\ref{RTeq}).

\section{ Bondi mass}
There are many possibilities how to compute a "mass" characterizing a given spacetime and its content. The definitions are built in such a way that they give expected results in situations where the correct answer seems obvious, like for the Schwarzschild solution with its mass parameter. When the spacetime geometry is suitable for a spacelike slicing the most natural one is either the ADM mass \cite{ADM} (asymptotic flatness is usually assumed for the construction) or the Komar mass \cite{Komar} (suitable for stationary spacetimes). But because of the standard formulation of the initial conditions for the Robinson-Trautman class (they are given on a null hypersurface) and the ensuing evolution in the retarded time direction we use the concept best suited to such a situation --- the Bondi mass \cite{Bondi}.
First, let us transform the original metric (\ref{ourmetric})  by the following change of variables (inspired by \cite{Podolsky-Svarc}) and one redefinition of a function
 \begin{equation}\label{trans}
 \rho=r\,\sqrt{U} \,\,\,\, , \,\,\, \chi=\frac{C_{0}}{\sqrt{U}}\,\,\,,\,\,\,\d 
\tilde{u}=\frac{\d u}{\sqrt{U}}\ .
 \end{equation}
 In terms of the new variables and the function $\chi$ the scalar field becomes 
 \begin{equation}\label{scalar-trans}
\varphi(\tilde{u},\rho)=\frac{1}{\sqrt{2}}\ln{\left\{ \frac{\rho- \chi( \tilde{u})}{\rho+ \chi( \tilde{u})}\right \}}
\end{equation}
(note the similarity with the potential of a finite rod in prolate ellipsoidal coordinates) and the metric simplifies into the following one
 \begin{equation}\label{general form}
 \d s^2 = -k(x,y)\,\d \tilde{u}^2-\,2\,\d \tilde{u}\,\d \rho +\,\frac{ \rho^2-{ 
\chi( \tilde{u})^2}}{{P(x,y)}^2}\,(\d y^{2} + \d x^{2})  ,
 \end{equation}
 which gives an easier interpretation for the special cases in the next section and is also suitable to investigate the global evolution of the energy content in our spacetime. On the other hand, the coordinate $\tilde{u}$ cannot be integrated in a closed analytic form. 

Now, we perform a conformal transformation of (\ref{general form}) so that the corresponding unphysical metric can be extended to the future null infinity $\mathcal{I}^{+}$ in a standard way. Since the vector $\partial_{\rho}$ is normal to $\mathcal{I}^{+}$ we select an appropriate conformal factor $\Omega=1/{\rho}$ and introduce a new coordinate $l=1/{\rho}$ for convenience. The unphysical metric $\d \tilde{s}^{2}=\Omega^{2} \d s^2$ then reads
\begin{equation}
  \d \tilde{s}^{2} = -k(x,y)l^{2}\,\d \tilde{u}^2+\,2\,\d \tilde{u}\,\d l +\,\frac{ 1-{\chi( \tilde{u})^2}l^{2}}{{P(x,y)}^2}\,(\d y^{2} + \d x^{2})  .
\end{equation} 
The coordinates of this metric represent the asymptotic Bondi coordinates near $\mathcal{I}^{+}$ and the tetrad adapted to $\mathcal{I}^{+}$ has the following form
\begin{eqnarray}
  \mathbf{l}&=&-\partial_{l}\ ,\nonumber\\
	\mathbf{k}&=&\partial_{ \tilde{u}}+\frac{k(x,y)\,l^{2}}{2}\,\partial_{l}\ ,\\
	\mathbf{m}&=&\frac{P(x,y)}{\sqrt{2( 1-{\chi^2}l^{2})}}(\partial_{x}+i\partial{y})\ .\nonumber
\end{eqnarray}

Since our spacetime contains a free massless scalar field we have to use a generalization of the standard covariant formula for calculation of the Bondi mass (based on a twistor equation) given in \cite{Scholtz} (note the sign change due to a different signature convention)
\begin{equation}
  M= \frac{1}{2\sqrt{\pi}}\oint \left[\Psi^{(0)}_{2}+2\Lambda^{(0)}+\sigma^{(0)} \dot{\bar{\sigma}}^{(0)}\right]\d S\ .
\end{equation}
In the above formula there appear the leading order terms of expansions of the standard Newman--Penrose quantities near the future null infinity. The integration is taken over a constant $u$ (or equivalently constant $\tilde{u}$) spatial sections of $\mathcal{I}^{+}$. Since our spacetime is shearfree the last term is identically zero while the first two combine to give
\begin{equation}
  M=-\frac{1}{2\sqrt{\pi}} \oint {\chi( \tilde{u})^2}_{,\tilde{u}} \,\d S = \frac{C_{0}^{2}U(u)_{,u}}{\sqrt{\pi}U(u)^{\frac{3}{2}}}\oint \d S\ .
\end{equation}
The surface area in the last expression is a finite positive constant and by inspecting the explicit form of $U(u)$ we can see that asymptotically the Bondi mass smoothly decreases to zero. Also, the Bondi mass is completely given by the scalar field and is identically zero when the scalar field is switched off by putting $C_{0}=0$. This confirms the previous result that our solution is related to the standard Robinson-Trautman geometry with vanishing mass parameter since otherwise the Bondi mass would contain additional contribution proportional to this parameter \cite{Tafel}. The natural interpretation of these results is that asymptotically the scalar field completely disappears being radiated away along $\partial_{\rho}$.

\section{Special cases}
Now we focus on special values of the parameters determining the general solution. Specifically, we consider the simplified form of metric (\ref{general form}) for the analysis.

\subsection{Case $C_{0}=0$}
If we assume $C_{0}$ is zero $(\chi=0)$ it means that the scalar field vanishes and 
the condition for $\omega$ (\ref{constraint}) (if it should remain finite) means that $\alpha$ should approach zero as well which 
leads to $\Delta k=0$. If we assume that the two-spaces spanned by $x,y$ are compact then the solution of the Laplace equation is necessarily a constant (the only harmonic functions on compact surfaces are constants). Since we assume the two-spaces to be regular its Gaussian curvature (determined by $k$) should be positive. Without loss of generality this constant can be chosen to be $1$ and we obtain a spherically symmetric situation, i.e.,
\[P^2 \rightarrow P_{0}^2 \,\,\,,\,\,\,\,k \rightarrow 1\ ,\]
where
\[P_{0}^2=1+\frac{1}{4}(x^2+y^2)\ .\]
It is convenient to choose $\d \Omega^{2}=\frac{(\d y^{2} + \d x^{2})}{P_{0}^2}$ so the metric would be
\[\d s^2 = -\,\d \tilde{u}^2-\,2\,\d \tilde{u}\,\d \rho +\rho^2 \,\d \Omega^{2},\]
which is obviously a flat spacetime and all the components of the Weyl spinor are zero. 

As shown in the section \ref{section-asymptotic} the final state of the asymptotic evolution corresponds to a flat solution as well so the case of a vanishing scalar field just considered is the future attractor for the general solution in the class.

\subsection{Case $\alpha=0$}
In this case, we have
\[\Delta k(x,y)={\alpha^2}=0\]
or in other words, as in the previous case, $k(x,y)$ is a constant $(k=1)$ and 
$P=P_{0}$. The metric functions $U$ and $h$ become (note that now $\omega=0$ from (\ref{constraint}))
\[U(u)=\gamma e^{\eta u}\, ,\ h(u,r)=\frac{\eta\,r}{2}\]
so the original metric can be written in the following form
\begin{equation}\label{Mlinear}
\d s^2 = -\left[\eta\,r+\frac{1}{U}\right]\,\d u^2-\,2\,\d u\,\d r +\left( 
Ur^2-\frac{C_{0}^2}{U}\right)\,\d \Omega^{2},
\end{equation}
and all the Weyl scalars are zero except $\Psi_{2}$,
\[\Psi_{2}=\frac{C_{0}^2}{3UR^{4}}\left(U_{,u}r-1\right).\]
Using the set of coordinates of the line element (\ref{general form}) the metric simplifies to 
this form
\begin{equation}\label{NML}
\d s^2 = -\d \tilde{u}^2-\,2\,\d \tilde{u}\,\d \rho + \left( \rho^2-{ 
\chi^2}\right)\,\d \Omega^{2}
\end{equation}
and the scalar field is retained in the form (\ref{scalar-trans}) with the function $\chi$ having a specific form derived below.
Since $\omega=0$, it is possible to solve the integral defining $\tilde{u}$ in the transformation (\ref{trans}) analytically, namely  
\begin{equation}\label{u-transf-DSS}
\tilde{u}=-\frac{2}{\eta\,\sqrt{\gamma\,e^{\eta u}}}+C\ ,
\end{equation}
where we can fix the constant of integration $C$ by demanding $u=0 \Rightarrow \tilde{u}=0$ to obtain
\[C=\frac{2}{\eta\,\sqrt{\gamma}}\, .\]
Now we can write $U$ explicitly in terms of $\tilde{u}$
\begin{equation}
U(\tilde{u})=\frac{\gamma}{\left(1-\frac{\eta \sqrt{\gamma}}{2} \, \tilde{u}\right)^2}\ .
\end{equation}
This expression has moreover a reasonable limit for $\eta \to 0$ due to our choice of the constant $C$.
In terms of the new variables the scalar field becomes 
\begin{equation}\label{scalar-field-DSS}
\varphi(\tilde{u},\rho)=\frac{1}{\sqrt{2}}\ln{\left\{ 
\frac{C\rho-{\chi_{0}}\,(C-\tilde{u})}{C\rho+ 
{\chi_{0}}\,({C}-\tilde{u})}\right \}}\ .
\end{equation}
where $\chi_{0}=\frac{C_{0}}{\sqrt{\gamma}}$.

The Ricci scalar and the Kretschmann invariant are giving the position of a curvature singularity
\begin{eqnarray}\label{position-sing}
RicciSc&=&\frac{2\chi(2\rho\chi_{,\tilde{u}}+\chi)}{(\rho^2-\chi^2)^2}=\\
&=&\frac{2\, C^{2}\chi_{0}^2 \,(\tilde{u}-C)\,\left[2\rho-(C-\tilde{u})\right]}{\left
[C^2 \rho^2-\chi_{0}^2\,(C-\tilde{u})^2\right]^2}\ ,\nonumber
\end{eqnarray}
(using a specific form of the solution)
\[Kretschmann=3(RicciSc)^2\ .\]
Obviously, the position given by the root of denominator changes linearly in $\tilde{u}$. If one looks for an apparent or a trapping horizon possibly covering the singularity one arrives at the following equation for the horizon hypersurface (derived from the condition for vanishing expansion of a congruence orthogonal to a spherically symmetric section of the horizon hypersurface)
\begin{equation}\label{horizon}
\rho_{h}=-(\chi^2)_{,\tilde{u}}\ ,
\end{equation}
for convenience it is possible to write $\chi$ in terms of $\tilde{u}$, namely
\[\chi(\tilde{u})=\frac{\chi_{0}}{C}(C-\tilde{u})\ ,\]
so \ref{horizon} would be
\begin{equation}
\rho_{h}=\frac{2\chi_{0}}{C}\chi(\tilde{u})\ .
\end{equation}
At the same time we know that the singularity is at $\rho_{sin}=\pm \chi(\tilde{u})$ using (\ref{position-sing}).

By a simple coordinate transformation the metric (\ref{NML}) and the scalar field (\ref{scalar-field-DSS}) can be shown to exactly correspond to the "nonstatic spherically symmetric massless scalar field" solution discussed by Roberts \cite{Roberts}. The presence of a horizon in this solution for certain values of parameters is briefly mentioned in \cite{Zhang} where more general spherically symmetric scalar fields with nontrivial potentials are discussed.

\begin{figure}[h]
\centering 
\includegraphics[trim = 15mm 140mm 80mm 20mm,scale=0.7]{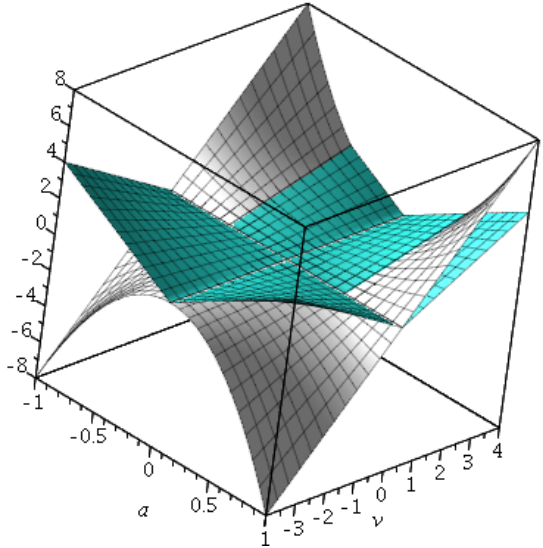}
\caption{Plot of the positions of the horizon (grey) and the singularity (green) for a range of the advanced time $v=C-\tilde{u}$ and the parameter $a=\chi_{0}/C$}
    \label{plot}
\end{figure}

Introducing a reparametrization of the retarded time $v=C-\tilde{u}$ (notice the reversal of the time direction) and an auxiliary parameter $a=\chi_{0}/C$ we can understand the relative positions of the singularity and the horizon by plotting them as in the Figure \ref{plot}. From the plot one sees that for the range of values $a=-1/2\ldots 1/2$ the singularity is permanently naked while for the rest of the values the horizon is always present above the singularity for positive times $v$. 

One can immediately recognize that the above time reversal corresponds to changing from the retarded to the advanced time. If one would not reverse the time orientation one would start (for $|a|>1/2$) with singularity covered by a large horizon at negative time which would gradually shrink and merge with the singularity at time zero.

Note that now the original asymptotic value of $u=\infty$ translates according to (\ref{u-transf-DSS}) into $\tilde{u}=C$ or $v=0$. There we immediately obtain $\chi=0$ and the scalar field (\ref{scalar-field-DSS}) vanishes as well. 

Interestingly, the Bondi mass now becomes proportional to the horizon position which indicates that in this spherically symmetric and dynamical case the Bondi mass (completely generated by the scalar field) plays the role similar to a variable mass in the Vaidya spacetime \cite{Vaidya} in determining the horizon position.

\subsection{Case $\omega=0$ and $\eta=0$} 
If we assume both $\omega$ and $\eta$ to be vanishing then $U(u)$ becomes a constant 
\[U=\gamma\]
and the whole geometry becomes obviously static and spherically symmetric.
In this case the line element (\ref{NML}) will simplify into
\begin{equation}\label{static-metric}
\d s^2 = -\d \widetilde{u}^2-\,2\,\d \widetilde{u}\,\d \widetilde{\rho} + \left( 
\widetilde{\rho}^2-{ \chi_{0}^2}\right)\,\d \Omega^{2},
\end{equation}
where $\widetilde{\rho}=r\,\sqrt{\gamma}$ and $\d \widetilde{u}=\frac{\d 
u}{\sqrt{\gamma}}$. Like in the previous case all the Weyl scalars are zero except 
$\Psi_{2}$ which becomes
\[\Psi_{2}=-\frac{\chi_{0}^2}{3\, \left( \widetilde{\rho}^2-{ 
\chi_{0}^2}\right)^2}\]
The scalar field is static as well 
\begin{equation}\label{static-field}
\varphi(\widetilde{\rho})=\frac{1}{\sqrt{2}}\ln{\left\{ 
\frac{\widetilde{\rho} -\chi_{0}}{\widetilde{\rho}+ \chi_{0}}\right \}}
\end{equation}
The Ricci scalar and the Kretschmann invariant are
\begin{equation}\label{RicciScalar}
RicciSc=\frac{2\,\chi_{0}^2}{\left(\widetilde{\rho}^2-\chi_{0}^2\right)^2}
\end{equation}
\[Kretschmann=3(RicciSc)^2\]
One can easily see that the singularity is naked in this case, either directly from the metric (\ref{static-metric}) or by looking for marginally trapped surfaces. The Bondi mass is now vanishing which might seem surprising at first but it is exactly in accordance with the observation made in the previous case that related the Bondi mass to the horizon position --- now the horizon is absent.

We can compare this static solution with the spherically symmetric static solution of Janis, Newmann and Winicour \cite{JNW1} in the coordinates given in \cite{JNW2, Boosting}
\begin{equation}\label{metric2}
\d s^{2}=-f(\tilde{R}) dt^{2}+\frac{1}{f( \tilde{R}) }\left \{{d\tilde{R}^{2}}+(\tilde{R}^{2}-M^{2}){d\Omega^{2}}\right \},
\end{equation}
in which 
\begin{equation}\label{function2}
f(\tilde{R})=\left [\frac {\tilde{R}-M}{\tilde{R}+M}\right]^{\frac{1}{\mu}}
\end{equation}
and the scalar field is $\phi={A}\,\ln [f(\tilde{R})]$ with the following relation between constants $\mu=\sqrt{1+2A^2}$. One immediately sees that in the limit $\mu \to \infty$ both the metric and the scalar field become identical (up to a trivial introduction of a null coordinate) to the static case given by (\ref{static-metric}) and (\ref{static-field}). 

Another connection to the previously studied spacetime can be found in the paper by Morris and Thorne \cite{Morris} studying traversable wormholes. Namely, the toy model of a wormhole spacetime proposed there can be obtained from (\ref{static-metric}) by a simple complex transformation of a constant $\chi_{0} \to i\chi_{0}$. This evidently means that the curvature scalars, e.g. (\ref{RicciScalar}), do not diverge anywhere and such a spacetime avoids the region with singularity by possessing a sphere with minimal areal radius which is nonzero. The scalar field becomes purely imaginary 
$$\varphi=\frac{i}{\sqrt{2}}\arg\left(\frac{\widetilde{\rho} - i\chi_{0}}{\widetilde{\rho}+ i\chi_{0}}\right)\ ,$$ so its stress energy tensor (being quadratic in $\varphi$) violates energy conditions as expected for a wormhole. Note that we consider the change $\chi_{0} \to i\chi_{0}$ as a parametric transition in our original Einstein-scalar field system of equations which means that we use the same definition for the stress energy of a scalar field as in \cite{Tahamtan-PRD-2015}, namely the standard stress energy of a real scalar field. Of course the stress energy tensor of a true complex scalar field does not lead to such a violation of the energy conditions needed here.

When ${\widetilde{\rho}}\to \infty$ the scalar field is still vanishing in this case. The metric (\ref{static-metric}) is also evidently asymptotically flat. But the area of the spherical surfaces $\widetilde{\rho}=const., u=const.$ grows quadratically with the coordinate $\widetilde{\rho}$ only far from the central region while close to the singularity $\widetilde{\rho}=\chi_{0}$ it grows just linearly.

\section{Imaginary scalar field}\label{wormhole}
Inspired by Morris and Thorne \cite{Morris} traversable wormholes and the simple relationship between their wormhole and our static solution \ref{static-metric}, we apply a complex transformation to appropriate constant $C_{0} \to i C_{0}$ in the general solution (\ref{ourmetric}). First, the scalar field becomes purely imaginary 
\begin{eqnarray}\label{imagi-scal}
\varphi(u,r)=\frac{1}{\sqrt{2}}\ln{\left\{ 
	\frac{U(u)r-i\,C_{0}}{U(u)r+i\,C_{0}}\right \}} \nonumber\\
\nonumber \\
=\frac{i}{\sqrt{2}}\arg \left( 
	\frac{U\,r-i\,C_{0}}{U\,r+i\,C_{0}}\right)
\end{eqnarray}
and the stress energy tensor (being quadratic in $\varphi$) violates all energy conditions. Second, the metric functions change accordingly
\begin{eqnarray}\label{imaginary-solution}
	R(u,r)&=&\sqrt{\frac{U(u)^{2}r^2+C_{0}^2}{U(u)}} \ , \nonumber\\
	U(u)&=&\gamma e^{-\frac{\alpha^2}{4\,C_{0}^2} u^2+\eta u}\ .
\end{eqnarray}
The metric sourced by a purely imaginary scalar field becomes  
\begin{eqnarray}\label{imag-metric}
\d s^2 &=& -\left[(-\frac{\alpha^2}{2\,C_{0}^2}\,u+\eta)\,r+\frac{k(x,y)}{U}\right]\,\d {u}^2-\,2\,\d {u}\,\d r   \nonumber \\
&&+\left(U\,r^2+\frac{C_{0}^2}{U}\right)\frac{(\d x^2+\d y^2)}{P(x,y)^2}
\end{eqnarray}
The Ricci scalar is
 \begin{equation}\label{RicciSc}
RicciSc=\frac{2C_{0}^2\,U\,\left(U_{,u}r-k\right)}{(U^{2}r^2+C_{0}^2)^{2}}
\end{equation}
and the Kretschmann invariant is again just its quadratic expression (using the specific form of the solution)
\[Kretschmann=3(RicciSc)^2\ .\]
This means that the curvature scalars do not diverge anywhere and curvature singularities are absent in this spacetime.

If we compute the expansion of the congruence associated with the vector field $\partial_{r}$
\begin{equation}\label{expansion-r}
  \Theta_{\partial_{r}} = \frac{2U^{2}r}{U^{2}r^{2}+C_{0}^{2}}
\end{equation}
we can see that by continuing the coordinate $r$ to negative values (note that $r=0$ is now neither a curvature singularity nor a coordinate one) we have a spacetime where the congruence $\partial_{r}$ changes sign at $r=0$. On the surface $r=0, u=const$ (which has a nonzero area) we not only have $\Theta_{\partial_{r}}=0$ but also $\partial_{r}\Theta_{\partial_{r}}>0$ so it is a genuine wormhole throat satisfying the flare-out condition \cite{Hochberg}. 

\begin{figure}[h]
\centering 
\includegraphics[trim = 15mm 140mm 80mm 20mm,scale=0.7]{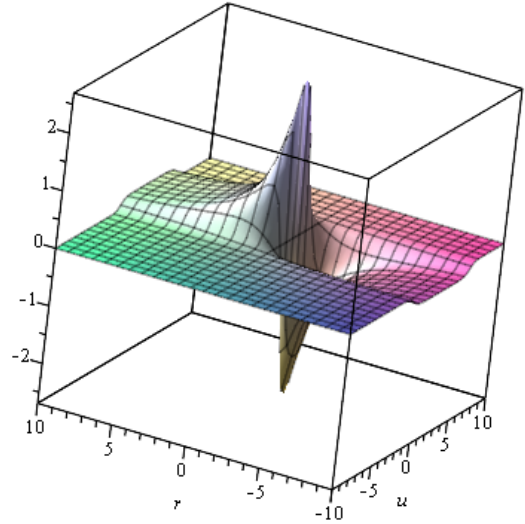}
\caption{Plot of the expansion of the congruence associated with the vector field $\partial_{r}$ in coordinates $r$ and $u$. All the arbitrary constants are set to $1$ for simplicity. The extremes are at the positions $u=2$, $r=\pm 1/\e$.}
    \label{fig2}
\end{figure}

By looking at the Figure \ref{fig2} one can recognize that the wormhole gets created only for a finite time close to the origin of coordinate $r$. To show that this behavior is general (if we consider $\alpha\neq 0$ and $C_{0}\neq 0$) one can easily compute that the expansion (\ref{expansion-r}) has two extremes (a positive maximum and a negative minimum) for $u=\frac{2C_{0}^{2}\eta}{\alpha^{2}}$ and $r=\pm\frac{C_{0}}{U}$. Combining this with the asymptotic behavior of the function $U$ (its explicit form (\ref{imaginary-solution}) shows that it vanishes for both $u\to \pm\infty$) and the value of the expansion (\ref{expansion-r}) for $r\to \pm\infty$ one immediately concludes that the throat structure analogous to that visualized in the Figure \ref{fig2} is generic.

The asymptotic form of the metric (\ref{imag-metric}) when $u$ goes to either positive or negative infinity is (note that now, unlike in the case of a real scalar field (\ref{solution}), the function $U$ asymptotically vanishes exponentially fast (\ref{imaginary-solution})),  
\begin{eqnarray}
	\d s^2 = -\left[\frac{k(x,y)}{U}\right]\,\d {u}^2-\,2\,\d {u}\,\d r+\left(\frac{C_{0}^2}{U\,P^2}\right)(\d x^2+\d y^2)\nonumber
\end{eqnarray}
and the scalar field \ref{imagi-scal} becomes a constant $-\frac{i\pi}{\sqrt{2}}$. The absence of any dependence on the coordinate $r$ in the two-dimensional metric defined on the subspace spanned by $x,y$ evidently means that the expansion of the congruence associated with the principal null vector $\partial_{r}$ is vanishing. So this limiting form of geometry belongs to the Kundt class (compare with (\ref{Kundt-metric}) for vanishing $W_{1}$ and $W_{2}$) which is a nonexpanding counterpart of the Robinson--Trautman family. 

The wormhole solution given above thus has genuine Robinson-Trautman behavior for finite times but asymptotically ($u\to \pm\infty$) transforms into a Kundt geometry. Specifically, it is related to a specific Kundt geometry coupled to a scalar waves (with $\Lambda=0$) discussed in \cite{ourKundt}.

The covariant (inverse trace) energy momentum tensor and its trace are zero asymptotically while the Weyl scalar $\Psi_{0}$ and the Ricci coefficient $\Phi_{00}\to \frac{-k(x,y)^2}{4 C_{0}^{2}}$ are nonzero. This behavior just means that certain tetrad projections are nonvanishing in the limit due to the behavior of the tetrad vectors (or in other words the metric). On the other hand, inspecting the Ricci scalar (\ref{RicciSc}) and the Kretschmann scalar in the asymptotic limit one can see that the mentioned divergences do not occur as a result of a strong curvature singularity presence. Nonzero $\Psi_{0}$ and $\Phi_{00}$ correspond to the Kundt-type gravitational and scalar waves both for the negative and positive infinite values of $u$ (see the form of the function $U$). One can interpret such a spacetime as containing a scalar wave (necessarily accompanied by a gravitational wave \cite{ourKundt}) coming from infinity and focusing to create a wormhole which is not stable and gets again radiated away in the form of waves of both fields.

 It is possible to rewrite the metric in the form similar to the real scalar field case (\ref{general form}) using the transformation (\ref{trans}), 
 \begin{equation}
 	\d s^2 = -k(x,y)\d \tilde{u}^2-\,2\,\d \tilde{u}\,\d \rho +\,\frac{\rho^2 +\chi^2}{P^2}\,(\d y^{2} + \d x^{2})  .
 \end{equation}

\section{Conclusion and final remarks}
We have presented additional properties of the Robinson--Trautman spacetime with a minimally coupled free 
scalar field. This important class of nonsymmetric and dynamical spacetimes was endowed with a scalar field source only recently. Using the asymptotic form of the solution and the Bondi mass we have shown that the scalar field is the only contribution to the energy of this solution and the energy content of the solution decreases to zero at the infinite retarded time. Accordingly the geometry itself becomes flat asymptotically. We have compared the asymptotic behavior of the vacuum and the scalar field solutions. The original investigation of the vacuum solution asymptotics required extensive analysis carried out mainly by P. Chru\'{s}ciel and led to the Schwarzschild solution (or its variants) as the final asymptotic state. 

Next, we have considered several special subcases of the general solution which all resulted in spherically symmetric situations. The $C_{0}=0$ case leads to a flat spacetime while $\alpha=0$ and $\omega=\eta=0$ cases retain the scalar field. In the $\alpha=0$ case both the scalar field and the geometry are dynamical while the $\omega=\eta=0$ case is completely static. We have shown that the dynamical case is similar to the Roberts solution while the static case corresponds to a limit of the Janis--Newmann--Winicour solution and is closely related to the simple version of the Morris--Thorne wormhole. This last correspondence led us to investigate a dynamical wormhole-type solution based on the general Robinson--Trautman spacetime with an imaginary scalar field. We showed that the wormhole throat appearance is generic and the asymptotics (both future and past one) is related to a subclass of the Kundt geometry with a scalar field. This provides a nontrivial connection between these two families of solutions to the Einstein equations on the level of a single spacetime.

\begin{acknowledgments}
This work was supported by the grant GA\v{C}R No. 14-37086G.
\end{acknowledgments}

\section*{APPENDIX}	
We present the Weyl scalars for the general solution (\ref{ourmetric},\ref{solution}). Note that in the original paper \cite{Tahamtan-PRD-2015} there are typos in the Weyl scalars presented there. Our preferred tetrad for determining the Weyl scalars of our solution 
is given by 
\begin{eqnarray}
	\mathbf{\tilde{l}}&=&\partial_{r}\nonumber\\
	\mathbf{\tilde{k}}&=&\partial_{u}-(h+K)\partial_{r}\\
	\mathbf{\tilde{m}}&=&\frac{P}{\sqrt{2}R}(\partial_{x}+i\partial{y})\nonumber
\end{eqnarray}
where $i$ is a complex unit. The Weyl spinor computed from this tetrad has only 
the following nonzero components
\begin{eqnarray}\label{Weyl}
	\Psi_{0}&=&-\frac{1}{4UR^{2}}\left[\{P^2(k_{,x}-ik_{,y})\}_{,x}-i\{P^2(k_{,x}-ik_{,y})\}_{,y}\right]\nonumber\\
	\Psi_{1}&=&\frac{\sqrt{2}PR_{,r}}{4UR^{2}}(k_{,x}-i\,k_{,y})\\
	\Psi_{2}&=&\frac{C_{0}^2}{3UR^{4}}\left(U_{,u}r-k\right) \nonumber
\end{eqnarray}
As correctly computed in \cite{Tahamtan-PRD-2015} the general algebraic type is II and in the special case of $k(x,y)=const > 0$ 
(constant positive Gaussian curvature of a compact two-space spanned by $x,y$) the 
algebraic type becomes D consistent with spherical symmetry. However, our family of solutions does 
not contain nontrivial type N radiative geometries that contain line singularities penetrating each wave surface.

\end{document}